\pdfoutput=1

\documentclass[aps,prb,twocolumn,superscriptaddress]{revtex4}
\usepackage{graphicx} 
\usepackage{dcolumn}
\usepackage{bm}

\begin{document}

\title{Ultrafast Generation of Fundamental and Multiple-order Phonon Excitations in Highly-Enriched (6,5) Single-Wall Carbon Nanotubes}
 
\author{Y.-S. Lim}
\affiliation{Department of Nano Science and Mechanical Engineering and Nanotechnology Research Center, Konkuk University, Chungju, Chungbuk 380-701, Republic of Korea}
\author{A. R. T. Nugraha}
\affiliation{Department of Physics, Tohoku University, Sendai 980-8578, Japan}
\author{S.-J. Cho}
\author{M.-Y. Noh}
\affiliation{Department of Nano Science and Mechanical Engineering and Nanotechnology Research Center, Konkuk University, Chungju, Chungbuk 380-701, Republic of Korea}
\author{E.-J. Yoon}
\affiliation{Department of Chemistry, POSTECH, Pohang 790-784, Republic of Korea}
\author{H. Liu}
\affiliation{Beijing National Laboratory for Condensed Matter Physics, Institute of Physics, Chinese Academy of Sciences, Beijing 100190, China}
\author{J.-H. Kim}
\affiliation{Department of Electrical and Computer Engineering, Rice University, Houston, TX 77005, USA}
\altaffiliation{Present address: Sungkyunkwan University, Jangan-gu, Suwon, Republic of Korea}
\author{H. Telg}
\author{E. H. H\'aroz}
\affiliation{Center for Integrated Nanotechnologies, Los Alamos National Laboratory, Los Alamos, NM 87545, USA}
\author{G. D. Sanders}
\affiliation{Department of Physics, University of Florida, Gainesville, FL 32611, USA}
\author{S.-H. Baik}
\affiliation{Quantum Optics Research Division, Korea Atomic Energy Research Institute, Daejeon 305-353, Republic of Korea}
\author{H. Kataura}
\affiliation{Nanosystem Research Institute, National Institute of Advanced Industrial Science and Technology, Tsukuba 305-8562, Japan}
\author{S. K. Doorn}
\affiliation{Center for Integrated Nanotechnologies, Los Alamos National Laboratory, Los Alamos, NM 87545, USA}
\author{C. J. Stanton}
\affiliation{Department of Physics, University of Florida, Gainesville, FL 32611, USA}
\author{R. Saito}
\affiliation{Department of Physics, Tohoku University, Sendai 980-8578, Japan}
\author{J. Kono}
\email[]{kono@rice.edu}
\thanks{corresponding author.}
\affiliation{Department of Electrical and Computer Engineering, Rice University, Houston, TX 77005, USA}
\affiliation{Department of Physics and Astronomy, Rice University, Houston, TX 77005, USA}
\affiliation{Department of Materials Science and NanoEngineering, Rice University, Houston, TX 77005, USA}
\author{T. Joo}
\email[]{thjoo@postech.ac.kr}
\thanks{corresponding author.}
\affiliation{Department of Chemistry, POSTECH, Pohang 790-784, Republic of Korea}

\date{\today}

\begin{abstract}
Using a macroscopic ensemble of highly-enriched (6,5) single-wall carbon nanotubes, combined with high signal-to-noise ratio, time-dependent differential transmission spectroscopy, we have generated vibrational modes in an ultrawide spectral range (10--3000~cm$^{-1}$).  A total of fourteen modes were clearly resolved and identified, including fundamental modes of A, E$_1$, and E$_2$ symmetries and their combinational modes involving two and three phonons.  Through comparison with CW Raman spectra as well as calculations based on an extended tight-binding model, we were able to identify all the observed peaks and determine the frequencies of the individual and combined modes.  We provide a full summary of phonon frequencies for (6,5) nanotubes that can serve as a basic reference with which to refine our understanding of nanotube phonon spectra as well as a testbed for new theoretical models.
\end{abstract}


\maketitle


\section{Introduction}

Recent advances in separation and sorting techniques have allowed researchers to prepare highly-enriched, single-chirality macroscopic ensembles of single-wall carbon nanotubes (SWCNTs).\cite{ZhengetAl03Science,ZhengetAl03NatMat,ArnoldetAl05NL,ArnoldetAl06NN,GhoshetAl10NatNano,LiuetAl11NatComm,TuetAl11JACS,KhripinetAl13JACS} This impressive progress is revolutionizing the field of carbon nanotube physics and chemistry, enabling detailed chirality-dependent spectroscopy studies that were previously impossible to perform.\cite{HarozetAl10ACS,DuqueetAl11ACS,HarozetAl11PRB,HarozetAl12JACS,HarozetAl13NS,ZhangetAl13NL}  While many of the features in prior studies were obscured due to the many different chiralities present in the samples, in highly-enriched samples these features are better resolved, allowing one to study the intrinsic behaviors of one-dimensional (1-D) electrons, phonons, and excitons in SWCNTs in far greater detail.  Moreover, the availability of highly-purified single-chirality SWCNTs opens up new possibilities for the development of optoelectronic devices\cite{DresselhausetAl01Book,AvourisetAl08NP,JorioetAl08Book,Leonard09Book,NanotetAl12AM} that are useful for optical communications, spectroscopy, imaging, and sensing.

In this paper, we focus on the properties of phonons in SWCNTs generated through ultrafast optical excitation on a macroscopic ensemble of highly-enriched (6,5) SWCNTs.\cite{LiuetAl11NatComm}  Raman spectroscopy has been an indispensable tool for characterizing and understanding the electronic and vibrational properties of graphite, SWCNTs, and graphene.\cite{SaitoetAl98Book,ReichetAl04Book,DresselhausetAl10NL,SaitoetAl11AP,JorioetAl11Book,FerrariBasko13NN}   Much accumulated knowledge exists on the strong Raman-active fundamental modes in SWCNTs such as the radial breathing mode (RBM), the D-mode, the G-mode, and the G$^{\prime}$ (or 2D) mode, observed through resonance Raman scattering spectroscopy.  However, in contrast to graphite and graphene, SWCNTs are expected to exhibit many other fundamental modes arising from the large number of 1-D phonon branches that arise from circumferential quantization (or zone folding).  At present, there is little experimental information available on these additional modes and their combinational modes, except some Raman studies of intermediate frequency modes (IFMs) in mixed-chirality samples.\cite{FantinietAl05PRB,ChouetAl05PRL,HtoonetAl05PRL,LuoetAl07PRB,LuoetAl08PRB}  In addition to the problem that they are obscured in chirality-mixed samples, they are also difficult to detect because they couple weakly to the photo-excited carriers.

We utilized some of the advantages of coherent phonon (CP) spectroscopy\cite{KimetAl13CP} to successfully observe and identify fourteen distinct phonon features in (6,5) SWCNTs.  The two most important advantages of CP spectroscopy for the present study are its excellent frequency resolution ($\sim$0.4~cm$^{-1}$ in the current case) and the ease with which low-frequency modes can be observed due to the absence of the Rayleigh scattering or photoluminescence background.  
We have observed symmetry-allowed acoustic modes (with E$_1$ and E$_2$ symmetries) that are consistent with our calculations based on the extended tight-binding model.  We also observed strong IFMs, which can be assigned to combinations between out-of-plane tangential optical (oTO) and acoustic phonons (E$_1$, E$_2$, and E$_1$+E$_2$ symmetries).  Furthermore, we observed combinational modes between A symmetry optical phonons, that is, the RBM and the LO mode, and acoustic phonons (E$_2$ and E$_1$+E$_2$ symmetries).  Finally, we provide a comprehensive list of phonon frequencies for (6,5) SWCNTs with assignments for both fundamental and combinational modes, based on a detailed comparison with our theoretical calculations of the full phonon dispersion curves for the (6,5) SWCNTs.

\section{METHODS}
\label{methods}

Single-chirality (6,5) SWCNTs with 4~mM concentration were micelle suspended by 0.5 wt \% DOC through the multicolumn gel chromatography isolation method.\cite{LiuetAl11NatComm}   

Pump-probe differential transmission measurements were performed on the solution in a quartz cell with an optical path length of 1~mm at room temperature. The light source for E$_{11}$ excitation was a home-built Ti:Sapphire laser equipped with a double-chirped mirror set (DCM 7, Venteon, Inc.), which delivers pulses as short as 8~fs with a bandwidth over 400~nm centered around 800~nm.  The probe beam was spectrally resolved by a set of bandpass filters with a bandwidth of 5~nm, prior to the photodetector.  The intensity ratio between the pump and probe beams was $\sim$7/2.  Compared to frequency-domain Raman spectroscopy, time-domain CP spectroscopy has several advantages: a high frequency resolution, the absence of Rayleigh scattering background at low frequencies, no interference by photoluminescence, and an ability to directly measure vibration dynamics in the time domain.  In a CP experiment, the high-frequency limit that can be detected depends on the pulse widths of the pump/probe beams; 10~fs pulses can generate and detect vibrational frequencies as high as 3000~cm$^{-1}$.  Furthermore, time-resolved CP spectra can directly investigate not only the decay dynamics of each active phonon, such as anharmonic decay processes of optical phonons, but spectrally-resolved CP spectra also provide information on excited-state phononic levels, accessible for the third order optical process involved in the pump/probe scheme, and initial phase change of phonon oscillations related to the bandgap modulations induced by the radial breathing motion.\cite{LimetAl06NL,KimetAl13CP}  

The phonon dispersions of the (6,5) SWCNT were obtained within an extended (long-range and symmetry-adapted) tight-binding model employing a specific force constant set.\cite{LietAl04SSC}  This approach takes into account the curvature effects of the SWCNT in the calculation of the force constant parameters derived from the graphene force constants. The graphene force constant parameters were calculated by fitting the phonon dispersion from inelastic X-ray scattering experiment.\cite{MaultzschetAl04PRL}

\section{RESULTS}

\begin{figure}
\begin{center}
\includegraphics[scale=0.6]{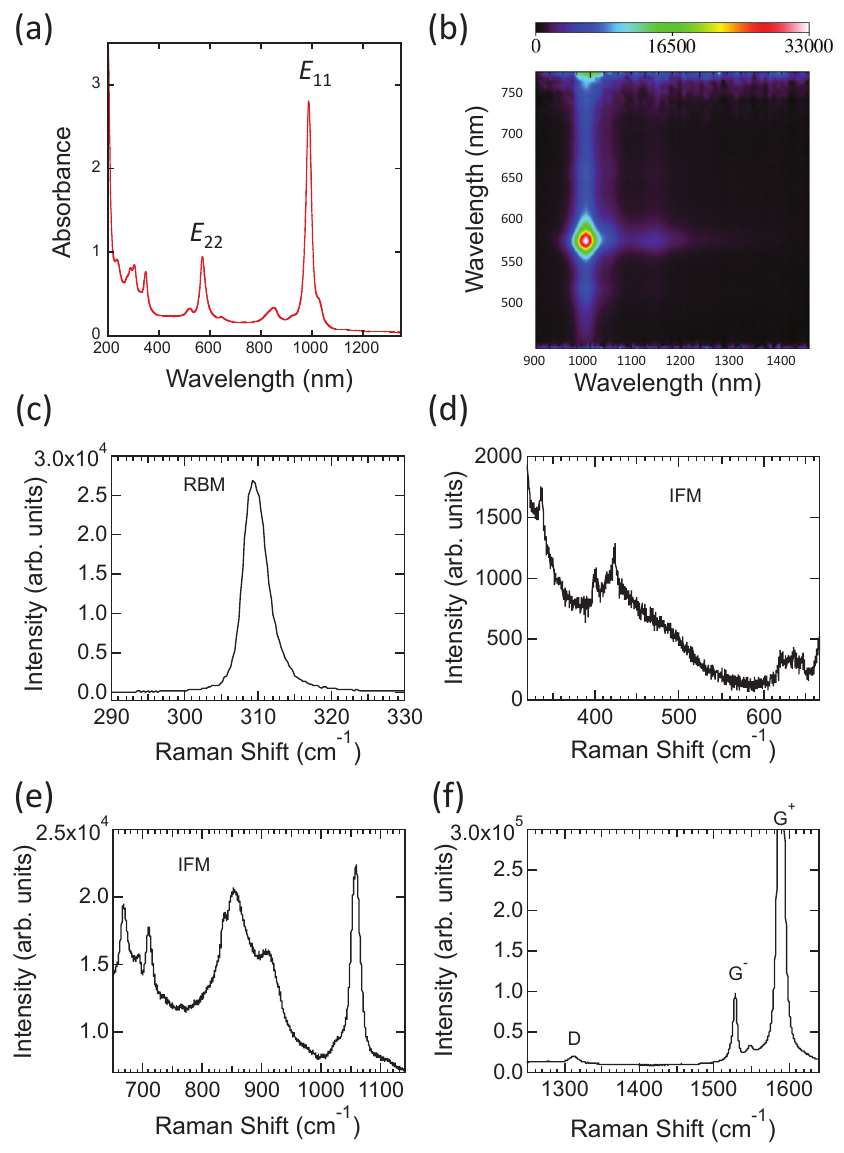}
\caption{(a)~Absorption and (b)~photoluminescence excitation spectra for the (6,5)-enriched SWCNT sample. (c)-(f)~Continuous-wave Raman spectra taken with an excitation wavelength of 561.3~nm for different frequency ranges from 290~cm$^{-1}$ to 1640~cm$^{-1}$.  Although the vertical scales are given in arbitrary units,  the relative scales are the same for (c)-(f).}
\label{sample}
\end{center}
\end{figure}

\begin{figure*}
\begin{center}
\includegraphics[scale=1]{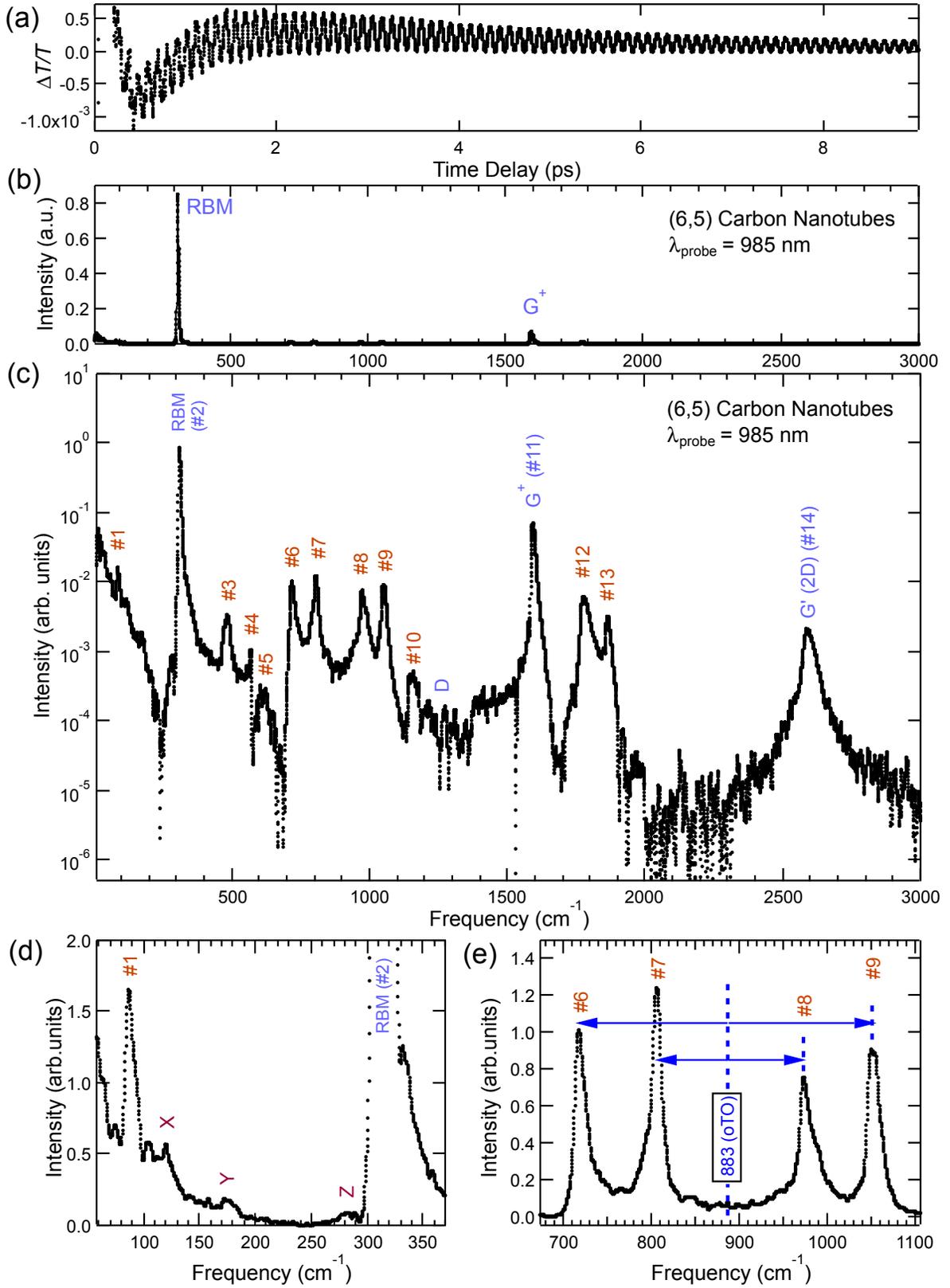}
\caption{(a)~Differential transmission trace at a probe wavelength of 975~nm for (6,5) single-wall carbon nanotubes showing ultrafast oscillations due to coherent phonons.  (b) and (c): Coherent phonon spectrum obtained through Fourier transformation of the data in (a) in linear and log scales, respectively, showing many phonon modes, labeled \#1-\#15.  (d)~The lowest frequency range, showing Peaks \#1-\#3, together with three less distinct peaks (X, Y, and Z).  (e)~Intermediate frequency modes, displaying symmetric pairs as sidebands of the oTO mode ($\sim$883~cm$^{-1}$).}
\label{exp}
\end{center}
\end{figure*}


Figure \ref{sample}a shows the absorption spectrum in the near-infrared, visible, and ultraviolet for the highly-enriched (6,5) SWCNT sample.  The largest absorption peaks are attributed to the lowest-energy interband resonances, E$_{11}$ and E$_{22}$, which occur at 985~nm (1.26~eV) and 575~nm (2.16~eV), respectively.  These values are in good agreement with those reported for other studies of (6,5) SWCNTs.\cite{NanotetAl12AM}  The other smaller features are due to the phonon sidebands of the E$_{11}$ and E$_{22}$ peaks as well as higher-order (E$_{33}$, E$_{44}$, etc.) interband transitions in the shortest wavelength region of the spectrum.  The narrow widths of the absorption lines attest to the high quality and purity of the (6,5) samples.  Figure~\ref{sample}b shows a photoluminescence excitation map for the sample, plotting the excitation wavelength on the vertical axis versus emission wavelength on the horizontal axis.  The strong spot observed at an excitation wavelength of 575~nm and an emission wavelength of 985~nm is consistent with the E$_{11}$ and E$_{22}$ peaks in the absorption spectrum.  No other spot is visible in the map, demonstrating the highly chirality-enriched nature of the sample.  Figures~\ref{sample}c-f show continuous-wave (CW) Raman scattering spectra, taken {\it via} E$_{22}$ resonance at an excitation wavelength of 561.3~nm, for different frequency ranges between 290~cm$^{-1}$ and 1640~cm$^{-1}$.  Although the vertical scales are given in arbitrary units,  the relative scales are the same for Figs.~\ref{sample}c-f.  In addition to the strong RBM (in c) and the D-mode and G-modes (in f), some weaker-intensity IFMs are observed (in d and e).  The frequencies of these features are summarized in the fourth column of Table~\ref{list}.

Figure \ref{exp}a shows a typical differential transmission trace in the time domain taken with $\sim$8~fs pump pulses with a bandwidth over 400~nm centered around 800~nm together with a probe beam with a 5-nm bandwidth centered at 975~nm.  The data exhibits ultrafast oscillations, containing various frequency components, the strongest of which are the RBM at $\sim$310~cm$^{-1}$ and the G$^+$ (LO) mode at $\sim$1590~cm$^{-1}$, as seen in the Fourier spectrum (Fig.~\ref{exp}b).  Figure~\ref{exp}c shows the same spectrum as \ref{exp}b but on a logarithmic scale in order to highlight a number of smaller-intensity modes.  We can clearly identify fourteen distinct features, labeled \#1-\#14 in Fig.~\ref{exp}c, the highest-frequency mode being the G$^{\prime}$ (or 2D) mode at $\sim$2590~cm$^{-1}$.  An expanded view of the lowest-frequency region of the spectrum is given in Fig.~\ref{exp}d, showing Peak \#1 at $\sim$85~cm$^{-1}$ and Peak \#2 (RBM) at $\sim$310~cm$^{-1}$.  In addition, there are three less distinct features at $\sim$120~cm$^{-1}$, $\sim$180~cm$^{-1}$, and $\sim$280~cm$^{-1}$, which we label `X', `Y', and `Z', respectively.  Finally, the IFMs are shown in Fig.~\ref{exp}e, displaying the pair-wise appearance of six peaks above and below a central frequency of $\sim$883~cm$^{-1}$.  The frequencies of the observed fourteen peaks with estimated error bars are listed in the second column of Table~\ref{list}, together with their respective assignments in the third column, as discussed below.  

\section{DISCUSSION}

\begin{figure*}
\begin{center}
\includegraphics[scale=0.55]{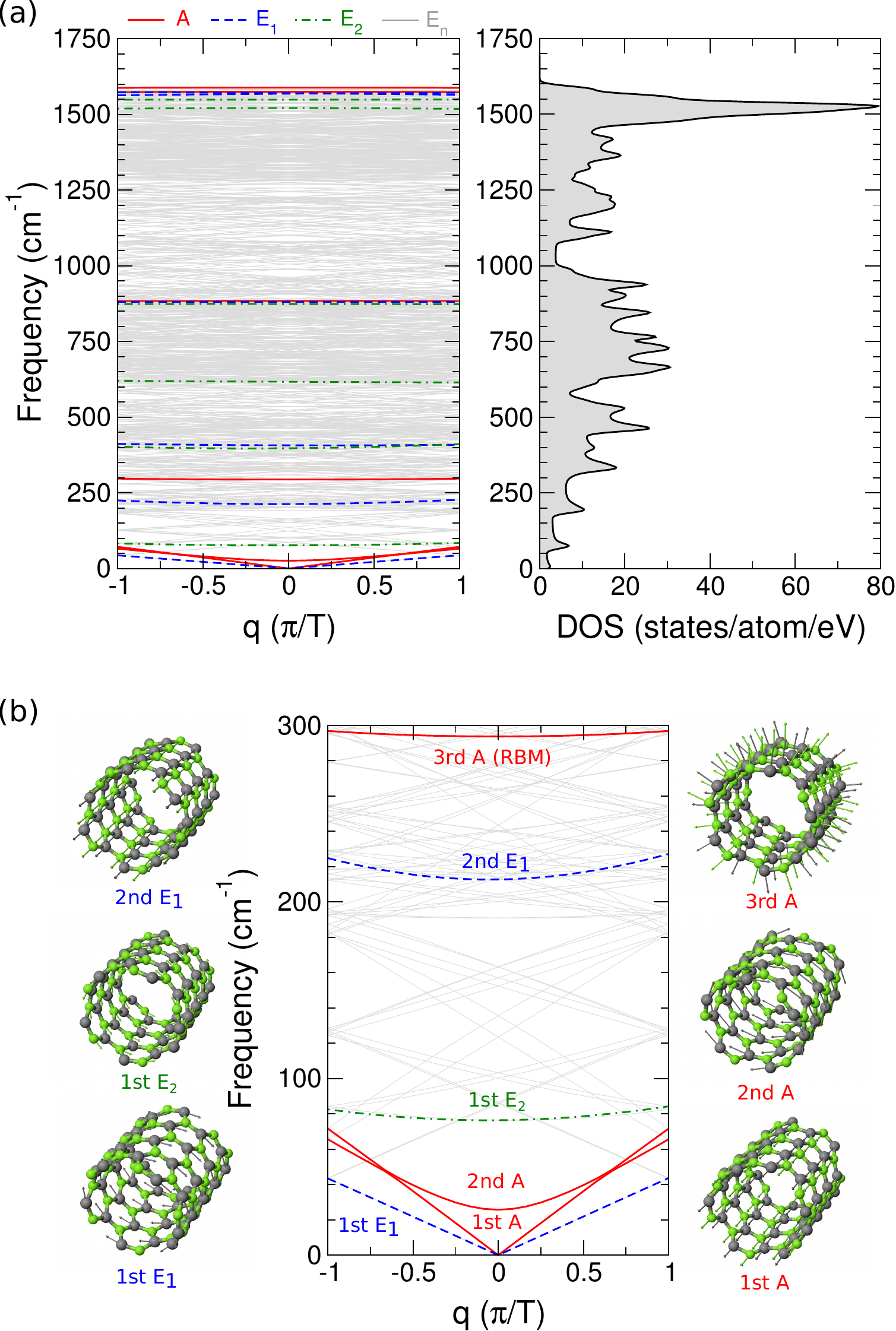}
\caption{(a)~Calculated phonon dispersion curves (left) and corresponding density of states (right) for possible phonon modes (A, E$_1$, E$_2$) for the (6,5) carbon nanotube, obtained by the extended tight-binding model.  Each symmetry mode is distinct from others by its color and has all the same six-fold phonon branches from the bottom.  (b)~Enlarged dispersion curve in the low-frequency region with pictures to display the corresponding eigenvector motions of carbon atoms.}
\label{theory}
\end{center}
\end{figure*}

In a SWCNT, vibrational modes with A, E$_1$, and E$_2$ symmetries are Raman active, and there are six modes in each symmetry.\cite{SaitoetAl98Book,DresselhausetAl10NL,JorioetAl11Book,ReichetAl04Book}  Among the phonon modes of the same order, the cutting line of the A mode is at the zone center of the graphene $k$-space while the cutting line of the E$_1$ (E$_2$) mode is the nearest (next nearest) to the zone center.  Among the six A symmetry modes, the RBM, iTO, and LO are expected to have non-zero electron-phonon matrix elements at the $\Gamma$ point ($q$ = 0), and two acoustic phonon modes --- the `twiston' mode (TW) and the LA mode --- are expected to have matrix elements at non-zero $q$.  The E$_1$ and E$_2$ symmetry modes have weaker electron-phonon coupling than the A symmetry mode.\cite{JiangetAl05PRB2}   Figure~\ref{theory}a shows the phonon dispersion curves and the density of states for the (6,5) SWCNT calculated using the extended tight-binding model.\cite{JiangetAl05PRB2}    It shows all possible E$_1$, E$_2$ and E$_n$ symmetries as well as A symmetry phonon branches. 
To investigate possible acoustic phonon modes at frequencies lower than the RBM, we enlarge the low frequency portion of the dispersion curve in Fig.~\ref{theory}b to display two branches for A and E$_1$ symmetry modes and one branch for the E$_2$ symmetry mode.  In order to provide more insight into each mode, phonon dynamical matrix eigenvectors are shown in Fig.~\ref{theory}b, whose directions identify these modes as iTA-like or LA-like oscillations. 
In the last column of Table 1, we list the $q$ = 0 frequencies of these fundamental modes and some combinational modes between them.  Although curvature effects for the force constants are taken into account in our model, the approximation begins to break down for nanotube diameters smaller than $\sim$0.8~nm since the long-range force constants are significantly affected by the curvature of the nanotube.  
For lower-frequency modes, the influence of the curvature effects is relatively small.

Based on the frequencies of the calculated phonon modes in Fig.~\ref{theory}, we can now discuss the origins of the observed fourteen peaks.  First, we attribute Peak \#1 at $\sim$85~cm$^{-1}$ to the E$_2$ (iTA-like) mode, whose predicted value is 76~cm$^{-1}$.   Given the uncertainty of $\pm$50~cm$^{-1}$ in the theoretically calculated values, which depends on the choice of force constants,\cite{LietAl04SSC} this is reasonable agreement.  
We also speculatively assign the weak and broader peak at $\sim$175~cm$^{-1}$ (Peak Y) to the E$_1$ (LA-like) mode, whose calculated value is 213~cm$^{-1}$.  
In addition, the weak peak near 280~cm$^{-1}$ (Peak Z), just below the RBM, could be due to a combinational mode between the 1$^{\rm st}$ E$_2$ (iTA-like, 76~cm$^{-1}$) and the 2$^{\rm nd}$ E$_1$ (LA-like, 213~cm$^{-1}$) modes; such a combinational mode between acoustic phonons has been seen in a graphite whisker\cite{TanetAl01PRB} and multi-wall carbon nanotubes\cite{TanetAl02PRB} and is believed to arise through a double-resonance process.\cite{SaitoetAl01PRL}   

As seen in Fig.~\ref{exp}e, four pronounced IFM peaks (Peaks \#6-\#9) appear in pairs, symmetrically separated from the center frequency at 883~cm$^{-1}$.  We interpret this central frequency as that of the oTO mode, which is predicted to have a frequency of 884~cm$^{-1}$, again in excellent agreement.  Then the two pairs of IFM peaks can be understood as the sum ($+$) and difference ($-$) frequencies, respectively, of the oTO mode and i) the 1$^{\rm st}$ E$_2$ mode (iTA-like, 76~cm$^{-1}$) and ii) the 2$^{\rm nd}$ E$_1$ mode (LA-like, 213~cm$^{-1}$).  In addition, the weaker IFM peaks at 623~cm$^{-1}$ (Peak \#5) and 1145~cm$^{-1}$ (Peak \#10) can also be the sum and difference, respectively, of the oTO mode and the sum of i) and ii) (three phonon combinations) although it is also likely that Peak \#5 is the RBM overtone as previously observed in Raman.\cite{LuoetAl08PRB}  Furthermore, there are two other IFM peaks, that is, Peaks \#3 and \#4.  Peak \#3 at 487~cm$^{-1}$ can be assigned to a combinational mode between the RBM (294~cm$^{-1}$) and the 2$^{\rm nd}$ E$_1$ mode (LA-like, 213~cm$^{-1}$). 
Peak \#4 at 574~cm$^{-1}$ can be assigned to a combination of three different symmetry modes, that is, the RBM (A$_1$, 294~cm$^{-1}$), the 2$^{\rm nd}$ E$_1$ mode (LA-like, 213~cm$^{-1}$), and the 1$^{\rm st}$ E$_2$ mode (iTA-like, 76~cm$^{-1}$). 


\begin{table}
  \caption{Experimentally measured and theoretically predicted frequencies of various coherent phonon (CP) modes for (6,5) single-wall carbon nanotubes, including both fundamental and combinational modes, together with assignments and CW Raman values.  Peak Number refers to the labels indicated for the observed CP peaks in Fig.~\ref{exp}c.  
  The values in the Theory column were calculated through the extended tight-binding model.\cite{JiangetAl05PRB2}  
  }
  \label{list}
  \begin{tabular}{|c|c||c|c|c|c|}
    \hline 
    Peak & CP & Assignment & Raman & Theory\\
    Number& (cm$^{-1}$) & & (cm$^{-1}$) & (cm$^{-1}$)\\
    \hline
    \hline
	\#1 & 86$\pm$3 &1$^{\rm st}$ E$_2$ (iTA) & & 76\\
	\hline
	& & 2$^{\rm nd}$ E$_1$ (LA) & & 213\\
	\hline
	\#2 & 309$\pm$2 & 3$^{\rm rd}$ A (RBM) & 309 & 294\\
	\hline
	 &  & 2$^{\rm nd}$ E$_2$ (LA-like) & & 397\\
	\hline
	&& 3$^{\rm rd}$ E$_1$ & 399 & 407\\
	
	&& 3$^{\rm rd}$ E$_1$ & 423 & 407\\
	\hline
	\#3 & 487$\pm$4 & RBM + 2$^{\rm nd}$ E$_1$ & & 507\\
	\hline
	\#4 & 574$\pm$4 &1$^{\rm st}$ E$_2$ + 2$^{\rm nd}$ E$_1$ + RBM & & 582\\
	\hline
	\#5 & 623$\pm$9 & RBM overtone  or & 621 & 588\\
	&& oTO $-$ 1$^{\rm st}$ E$_2$ $-$ 2$^{\rm nd}$ E$_1$ &  & (595)\\
	\hline
	&&3$^{\rm rd}$ E$_2$ & & 616\\
	\hline
	&& & 668 & \\
	\hline
	&& & 692 & \\
	\hline
	\#6 & 718$\pm$1 & oTO $-$ 2$^{\rm nd}$ E$_1$ & 710 & 671\\
	\hline
	\#7 & 807$\pm$5 & oTO $-$ 1$^{\rm st}$ E$_2$ & & 807\\
	\hline
	&& & 837 & \\
	\hline
	&& & 853 & \\
	\hline
	& 883$\pm$12 & 4$^{\rm th}$ A (oTO) & & 884\\
	\hline
	&& 4$^{\rm th}$ E$_2$ & & 874\\
	\hline
	&& 4$^{\rm th}$ E$_1$ & & 881\\
	\hline
	&& & 908 & \\
	\hline
	\#8 & 970$\pm$6 & oTO + 1$^{\rm st}$ E$_2$ & & 960\\
	\hline
	&& & 1027 & \\
	\hline
	\#9 & 1054$\pm$3 & oTO + 2$^{\rm nd}$ E$_1$ & 1057 & 1096\\
	\hline
	\#10 & 1145$\pm$6 & oTO + 1$^{\rm st}$ E$_2$ + 2$^{\rm nd}$ E$_1$ & & 1173\\
	\hline
	&& D& 1312 & 1338\\
	\hline
	&& 5$^{\rm th}$ E$_2$ & & 1521\\
	\hline
	&& 5$^{\rm th}$ E$_1$ & & 1568\\
	\hline
	&& 5$^{\rm th}$ A (iTO or G$^{-}$) & 1528 & 1575\\
	\hline
	\#11 & 1588$\pm$4 & 6$^{\rm th}$ A (LO or G$^+$) & 1589 & 1588\\
	\hline
	&& 6$^{\rm th}$ E$_1$ & & 1570\\
	\hline
	&& 6$^{\rm th}$ E$_2$ & & 1548\\
	\hline
	\#12 & 1764$\pm$4 & LO + 2$^{\rm nd}$ E$_1$ & & 1801\\
	\hline
	\#13 & 1866$\pm$4 & LO + 1$^{\rm st}$ E$_2$ + 2$^{\rm nd}$ E$_1$ & & 1877\\ 
	\hline
	\#14 & 2591$\pm$4 & 2D (or G$^{\prime}$) & & 2618\\
	\hline
  \end{tabular}
\end{table}


Previous Raman measurements on chirality-mixed SWCNT samples detected some IFMs in the 600-1100~cm$^{-1}$ frequency range excited in the E$_{33}$/E$_{44}$ region\cite{FantinietAl05PRB} as well as in the 370-480~cm$^{-1}$ frequency range excited in the E$_{11}$/E$_{22}$ region.\cite{LuoetAl07PRB,LuoetAl08PRB}  Fantini {\it et al.}~observed only two peaks associated with the oTO mode combined with acoustic phonon modes.\cite{FantinietAl05PRB}  The so-called `step-wise' dispersive behaviors were observed in these studies because the IFMs have a strong chirality- and diameter-dependence, which jumps when the excitation photon energy goes from one 2$n+m$ family to another.  Kilina and co-workers\cite{KilinaetAl08PNAS} suggested some connection between the resonances for these particular IFMs and cross-polarized transitions, which may be reasonable with the E-symmetry character of these modes.  However, in all these previous studies on chirality-mixed ensembles, the analysis and ultimate ($n$,$m$) assignment of specific IFMs were model-dependent, and thus, the determined IFM frequencies depended on the assumsion that the ($n$,$m$) assignments were correct.  On the contrary, the present work on IFM coherent phonons represents a single ($n$,$m$) probe of IFM behavior, providing model-independent frequencies.  

Lastly, it is interesting to note some distinct differences between Raman and CP spectra.  First, Raman observed a clear D-peak at $\sim$1312~cm$^{-1}$ (see \ref{sample}f) whereas it was absent in the CP spectra (see Fig.~\ref{exp}c). Second, a clear G$^-$ peak was observed at $\sim$1528~cm$^{-1}$ in Raman (see Fig.~\ref{sample}f) whereas it was not clearly observed in CP (see Fig.~\ref{exp}c).  We are currently modeling the CP generation and detection processes for various phonon modes, including the D and G$^-$ modes, to compare their expected intensities with the experimental observations.  Furthermore, in Raman we see an IFM at $\sim$399~cm$^{-1}$, $\sim$420~cm$^{-1}$, $\sim$668~cm$^{-1}$, $\sim$692~cm$^{-1}$, $\sim$837~cm$^{-1}$, $\sim$853~cm$^{-1}$, $\sim$908~cm$^{-1}$, and $\sim$1027~cm$^{-1}$ (see Figs.~\ref{sample}d-e and Table~\ref{list}), which do not agree with any of the clearly observed CP peaks.  It should be noted that any short-lived phonon mode cannot be detected in CP spectroscopy because they are significantly affected by the fast and strong electronic pump-probe signal (especially important in SWCNTs).  In our CP analysis, we did not use any data before $\sim$125~fs, and therefore, if the peak width exceeds $\sim$80~cm$^{-1}$, the peak is not recorded.  This is a possible explanation as to why CP does not show the broad peaks at 853~cm$^{-1}$ and 908~cm$^{-1}$ observed in Raman.  In addition, the phonon generation mechanism by photo-excited carriers (exciton-phonon interaction in the case of SWCNTs) should be common for Raman spectroscopy and ultrafast pump-probe spectroscopy.  However, in the case of Raman spectroscopy, we measure the photon energy of the scattered light while in the case of pump-probe spectroscopy we measure the reflectance or transmittance of the probe beam.  The optical response of the material to the probe beam in the presence of the phonons may not always be the same as that for the Raman spectra.  The modifications of the probe reflectance and transmittance should depend on phonon vibration amplitudes and directions. A further investigation is under way.

\section{CONCLUSION}

In conclusion, by combining the excellent signal-to-noise ratio and spectral resolution of coherent phonon spectroscopy with a highly-enriched, single-chirality (6,5) single-wall carbon nanotube sample,  we have detected previously unobserved vibrational modes.  The combination of a modern SWCNT separation technique with advanced ultrafast optical spectroscopy allowed direct observations of several phonon branches of A, E$_1$, and E$_2$ symmetries and their combinations. 
We demonstrated that there is strong coupling between different phonon modes over a wide frequency range. 
By presenting a detailed compendium of phonons for single-chirality single-wall carbon nanotubes, this work serves as a reference with which to refine our understanding of the dynamics and interactions of one-dimensional phonons and electrons and as a testbed for new theoretical models for more precise and accurate calculations. 
Furthermore, the present analysis shows that there is a clear discrepancy for some phonon modes between Raman and pump probe measurements, whose reasons remain to be explained.

\section*{Acknowledgements}

This work was supported in part by the National Research Foundation of Korea (NRF) grant funded by the Korea government (MEST) (2013R1A1A2006659, 2010-022691). TJ acknowledges the support by the National Research Foundation of Korea (NRF) grant funded by the Korea government (MSIP) (2007-0056330) and the Global Research Laboratory Program (2009-00439). JK acknowledges support by the Department of Energy (through Grant No.~DE-FG02-06ER46308), the National Science Foundation (through Grants No.~OISE-0968405), and the Robert A. Welch Foundation (through Grant No.~C-1509).  GDS and CJS acknowledge support from the National Science Foundation through grants OISE-0968405 and DMR-1105437.  HL acknowledges support by the ``100 talents project'' of CAS and the recruitment program of global youth experts.  HK acknowledges support by JSPS KAKENHI Grant Number 25220602.  RS acknowledges support by KAKENHI (Nos.~25286005 and 25107005).  ARTN acknowledges the support by the JSPS Research Fellowship for Young Scientists (201303921).  Raman spectra were acquired at the Center for Integrated Nanotechnology, a U.S. Department of Energy, Office of Basic Energy Sciences user facility.  EHH, HT, and SKD acknowledge partial support from the LANL LDRD program.



\end{document}